\documentclass[letterpaper,11pt]{article}
\usepackage{jhepmod} 

\def\la{\langle}
\def\ra{\rangle}

\def\be{\begin{equation}}
\def\ee{\end{equation}}
\def\ben{\begin{eqnarray}}
\def\een{\end{eqnarray}}
\def\nn{\nonumber}
\def\oh{\bf \hat\Omega}

\def\bk{{\bf k}}

\def\bk{{\bf k}}

\def\bl{{\bf l}}

\def\2p{{(2\pi)^2}}

\def\bl{{\bf l}}
\def\be{\begin{equation}}
\def\ee{\end{equation}}
\def\beq{\begin{equation}}
\def\eeq{\end{equation}}
\def\ben{\begin{eqnarray}}
\def\een{\end{eqnarray}}
\def\bes{\begin{subequations}}
\def\ees{\end{subequations}}

\def\oh{{\hat\Omega}}
\def\nn{{\nonumber}}

\newcommand{\beqa}{\begin{eqnarray}}
\newcommand{\eeqa}{\end{eqnarray}}

\def\ikap0{{\cal J}_{\theta_0}(r)}

\def\one1{\langle \kappa_{(i)}\kappa_{(j)} \rangle}
\def\one{{[\bar \xi^{(ij)}]}}

\def\ba{\begin{eqnarray}}
\def\ea{\end{eqnarray}}

\def\bk{{\bf k}}

\def\bk{{\bf k}}

\def\bl{{\bf l}}

\def\2p{{(2\pi)^2}}

\def\bl{{\bf l}}

\def\be{\begin{equation}}
\def\ee{\end{equation}}

\def\beq{\begin{equation}}
\def\eeq{\end{equation}}

\def\ben{\begin{eqnarray}}
\def\een{\end{eqnarray}}

\def\oh{{\hat\Omega}}
\def\nn{{\nonumber}}

\def\bk{{\bf k}}

\def\bl{{\bf l}}

\def\2p{{(2\pi)^2}}

\def\bl{{\bf l}}

\def\bl{{\bf{l}}}

\def\ep{{\boldsymbol\epsilon}}
\def\th{{\boldsymbol\theta}}

\usepackage{booktabs}
\usepackage[english]{babel}
\usepackage{amsmath,amssymb,amsbsy,amstext, amsthm, simplewick, amsfonts}
\usepackage{hyperref}
\usepackage{graphicx}
\usepackage{subcaption}

\usepackage{siunitx}
\usepackage{upgreek}
\usepackage{framed}
\usepackage{wrapfig}
\usepackage{multirow}
\usepackage{bbm}
\usepackage[svgnames,dvipsnames,x11names]{xcolor}
\usepackage[utf8x]{inputenc}
\usepackage{selinput}
\usepackage{bm}
\usepackage{float}
\usepackage{yfonts}
\usepackage{mathtools}
\usepackage{enumitem}   
\usepackage{longtable}
\usepackage{tabularx}
\usepackage{amsmath}

\usepackage{xcolor}
\usepackage{soul}

\usepackage{xcolor}

\usepackage{tikz}
\usetikzlibrary{arrows,chains,matrix,positioning,scopes}
\usepackage{tikz-cd}
\usetikzlibrary{matrix, calc, arrows}

\def\sraise{\;\raise1.0pt\hbox{$'$}\hskip-6pt\partial}
\def\slower{\;\overline{\raise1.0pt\hbox{$'$}\hskip-6pt\partial}}

\usepackage{colortbl}

\makeatletter
\newlength{\apb@width}
\newcommand{\autoparbox}[2][c]{\settowidth{\apb@width}{#2}\parbox[#1]{\apb@width}{#2}}

\makeatother

\def\0{{\boldsymbol{0}}}

\def\nn{\nonumber\\}

\def\beq{\begin{equation}}
\def\eeq{\end{equation}}
\def\be{\begin{equation}}
\def\ee{\end{equation}}

\usepackage[framemethod=default]{mdframed}
\newmdenv[skipabove=7pt,
skipbelow=7pt,
rightline=false,
leftline=false,
topline=false,
bottomline=false,
backgroundcolor=gray!10,
linecolor=gray,
innerleftmargin=5pt,
innerrightmargin=5pt,
innertopmargin=5pt,
innerbottommargin=5pt,
leftmargin=0cm,
rightmargin=0cm,
linewidth=4pt]{eBox}

\title{A New Estimator for Phase Statistics}
\author{D. Munshi$^a$,  R. Takahashi$^b$, 
  J. D. McEwen$^c$, T. D. Kitching$^d$, F. R. Bouchet$^{e,f}$ \\} 
\affiliation{$^{a,c,d}$Mullard Space Science Laboratory, University College London, 
  Holmbury St Mary, Dorking, Surrey RH5 6NT, UK;
$^b$Faculty of Science and Technology, Hirosaki University, 3 Bunkyo-cho, Hirosaki, Aomori 036-8561, Japan,\;\;
$^{e}$Institut d’Astrophysique de Paris, UMR 7095,\;\; $^{f}$CNRS Sorbonne Universit, 98 bis Boulevard Arago, F-75014 Paris, France}

\emailAdd{$^a$D.Munshi@ucl.ac.uk, $^b$takahasi@hirosaki-u.ac.jp, 
  $^c$Jason.McEwen@ucl.ac.uk, $^d$t.kitching@ucl.ac.uk, $^e$bouchet@iap.fr}
\abstract{We introduce a novel statistic to probe the statistics of phases of Fourier modes
  in two-dimensions (2D) for weak lensing convergence field $\kappa$.
  This statistic contains completely independent information compared to that contained in observed power spectrum.
  We compare our results against state-of-the-art numerical simulations as a function
  of source redshift and find good agreement with theoretical predictions.
  We show that our estimator
  can achieve better signal-to-noise compared to the commonly employed
  statistics known as the line correlation function (LCF). Being a two-point
  statistics, our estimator is also easy to implement in the presence of complicated noise and mask,
  and can also be generalised to higher-order.
  While applying this estimator for the study of lensed CMB maps,
  we show that it is important to include post-Born corrections in the study of statistics of phase.}
\begin{document}

\maketitle
\flushbottom

\section{Introduction}
\label{sec:into}
The weak lensing surveys which include 
 Canada-France-Hawaii Telescope{(CFHTLS)}\footnote{\href{http://www.cfht.hawai.edu/Sciences/CFHLS/}{\tt http://www.cfht.hawai.edu/Sciences/CFHLS}},
{PAN-STARRS}\footnote{\href{http://pan-starrs.ifa.hawai.edu/}{\tt http://pan-starrs.ifa.hawai.edu/}},
Dark Energy Surveys (DES)\footnote{\href{https://www.darkenergysurvey.org/}{\tt https://www.darkenergysurvey.org/}}\citep{DES},
Prime Focus Spectrograph\footnote{\href{http://pfs.ipmu.jp}{\tt http://pfs.ipmu.jp}},
{WiggleZ}\footnote{\href{http://wigglez.swin.edu.au/}{\tt http://wigglez.swin.edu.au/}}\citep{WiggleZ},
{BOSS}\footnote{\href{http://www.sdss3.org/surveys/boss.php}{\tt http://www.sdss3.org/surveys/boss.php}}\citep{SDSSIII}, {KiDS}\citep{KIDS} 
and {Subaru Hypersuprimecam survey}\footnote{\href{http://www.naoj.org/Projects/HSC/index.html}{\tt http://www.naoj.org/Projects/HSC/index.html}}(HSC) \citep{HSC}
are already providing important cosmological insights.
The future large scale structure (LSS) surveys 
{\it Euclid}\footnote{\href{http://sci.esa.int/euclid/}{\tt http://sci.esa.int/euclid/}}\citep{Euclid},
{Rubin Observatory}\footnote{\href{http://www.lsst.org/llst home.shtml}{\tt {http://www.lsst.org/llst home.shtml}}}\citep{LSST_Tyson} and
Roman Space Telescope\footnote{\href{https://roman.gsfc.nasa.gov/}{https://roman.gsfc.nasa.gov/}}
list weak lensing as their main science driver and are expected to
take us beyond the standard model of cosmology \citep{Planck1} by answering some of
the most profound questions regarding the nature of dark matter and
dark energy (equivalently the modified theories of gravity)\citep{MG1,MG2}
and nature of neutrino mass hierarchy \citep{nu}.

Weak lensing observations target the low-redshift
universe and small scales where the perturbations are in the nonlinear regime and
statistics are non-Gaussian \citep{Munshi_Review}. Many different estimators exist which probe the
higher-order statistics of weak lensing maps \citep{Carbone}. These include
the well known real-space one-point statistics such as the
cumulants \citep{Barber1} or their two-point correlators also known as
the cumulant correlators as well as the associated PDF \citep{Uhlemann1} and
the peak-count statistics \citep{peak_count}.
In the harmonic domain 
the estimators such as the Skew-Spectrum\citep{skew},
Integrated Bispectrum \citep{Integrated}
kurt-spectra \citep{kurt}, morphological estimator \citep{morph},
integrated tripsectrum \citep{IT},
Betti number \citep{Betti},
extreme value statistics \citep{EVS}, position-dependent PDF \citep{pPDF},
density split statistics \citep{split}, response function formalism \citep{response}
estimators for shapes of the lensing bispectrum \citep{shape} are some of the statistical estimators
and formalism recently considered by various authors in the context of understanding
cosmological statistics in general and weak lensing in particular.

The gravitational clustering in the quasilinear and nonlinear regime
generates coupling of Fourier modes that results in correlation of their phases \citep{Bernard_Review}.
The power spectrum does not contain any phase information \citep{Phase}. Notable initial work 
on statistical evolution of phases of the Fourier modes in gravitational
clustering in the quasilinear regime can be found in  \citep{Matsubara1,Matsubara2}.
These studies relied on a perturbative framework and were tested against numerical simulations \citep{Hikage}.
A universal behavior in evolution of phases of the
Fourier modes in the nonlinear regime was reported in \citep{Watts_Coles_Melott}.

In recent years an estimator known also as the line correlation function (LCF) to
measure for three-point (third-order) phase correlations was introduced in \citep{Wol}
(also see \citep{Obr}). This was used in many different contexts.
The possibility of improving the cosmological constraints using phase correlations
were studied in \citep{Alex,Ali}.
In the context of redshift-space distortions this was used in \citep{Obreschkow2}. 
The growth rate of perturbation were probed in \citep{Byun}.
One of the motivation in this paper is to introduce the third-order
phase correlation functions for projected surveys in general and weak lensing
surveys in particular. We introduce third-order estimators using both two-
and three-point statistics.

We note here in passing that in addition to the summary statistics listed above,
in recent years many novel techniques have gained popularity. These
include Bayesian hierarchical modelling, likelihood-free or forward modelling approaches.
However, it is important to realise that many of these methods often
rely on simulations that are based on lognormal approximation or high-order
Lagrangian theories and are only approximate compared to accurate ray tracing simulations
which can be rather expensive (however, see also \citep{likefree1,likefree2}).

At leading order the LCF takes contributions from the bispectrum.
However, at smaller
separations, it also take contribution from higher-order statistics.
The perturbative treatment breaks down at smaller separation.
The LCF encodes information that is highly complementary to that
contained in the power spectrum. Next, we will
introduce a two-point statistics also known as cumulant-correlator
which can probe phase bispectrum with a higher signal-to-noise.

The LCF has also been employed to distinguish various morphological types of collapsed objects \citep{Obr}.
In addition to the LCF other real space triangular configurations (TCF) have been
considered for the study of third-order phase
statistics, e.g., \citep{Pritchard} employed TCF to probe the characteristic scale of
ionized regions during the epoch of reionization from $21$cm interferometric observations.

This paper is organised as follows.
 In \textsection\ref{WL} we introduce the
 weak lensing bispectrum.
 In \textsection\ref{sec:line} we briefly review the modelling of the
 third-order phase statistics.
 In \textsection\ref{sec:disc} we discuss our results.
 The conclusions are drawn in \textsection\ref{sec:conclu}.
 
\section{Weak Lensing Bispectrum}
\label{WL}
%
The projected weak lensing convergence $\kappa$ is a line-of-sight integration of the
underlying three-dimensional (3D) cosmological density contrast $\delta$. The $\kappa({\th})$
at a position $\th$ can be expressed as follows:
\ben
\kappa({\th}) = \int_0^{r_s} dr \; \omega(r)\; \delta(r,{\th}); \quad
w(r) = {3 \Omega_{\rm M} H_0^2\over 2 c^2 a} {d_A(r-r_s) d_A(r) \over  d_A(r_s)}.
\een
Here $d_A(r)$ is the comoving angular diameter distance at a comoving distance $r$.
The kernel $w(r)$ encodes geometrical dependence;
$a$ is the scale factor, $H_0$ is the hubble constant and $\Omega_M$ is the
cosmological density parameter. $d_A(r)$ and $d_A(r_s)$
are comoving angular diameter distances at a comoving
distances $r$ and $r_s$.  We have assumed all sources to be at a single
source plane at a distance $r_s$.

The power spectrum $P^{\kappa}(l)$ and the bispectrum $B^{\kappa}({\bl}_{1},{\bl}_{2},{\bl}_{3})$
for the convergence maps are defined through the following expression :
\bes\ben
&& \la\kappa({\bl}_{1})\kappa({\bl}_{2})\ra =  
(2\pi)^{2}\delta_{\rm 2D}({\bl}_{1}+{\bl}_{2})P^{\kappa}(l_{1}); \quad |\bl_i| = l_i
\label{eq:flat_sky_power}\\
&& \la\kappa({\bl}_{1})\kappa({\bl}_{2})\kappa({\bl}_{3})\ra 
= (2\pi)^{2}\delta_{\rm 2D}({\bl}_{1}+{\bl}_{2}+{\bl}_{3})B^{\kappa}({\bl}_{1},{\bl}_{2},{\bl}_{3}).
\label{eq:flat_sky_bispec}
\een\ees
The two dimensional Fourier transform of $\kappa(\th)$ is denoted as $\kappa({\bl})$ with
${\bf l}$ being the two-dimensional wave vector.
we denote the two-dimensional Dirac delta function as $\delta_{\rm 2D}$.
We will use $P_\delta$ and $B_\delta$ to denote the power spectrum and bispectrum respectively
of the underlying three dimensional cosmological density contrast $\delta$.
In the tree level standard perturbation  theory the bispectrum $B_{\delta}(\bk_1,\bk_2,\bk_3)$
can be expressed in terms of the kernel $F_2(\bk_1,\bk_2)$ and power spectrum as follows \citep{Bernard_Review}:
\bes\ben
&& B^{\rm PT}_{\delta}(\bk_1,\bk_2,\bk_3) = P^{\rm L}_{\delta}(k_1)P^{\rm L}_{\delta}(k_2) F_2(\bk_1,\bk_2) + {\rm cyc.perm.}\\
&& F_2(\bk_1,\bl_2) = {5 \over 7} +
{1 \over 2} \left [ {k_1 \over k_2} +{k_2 \over k_1} \right ]\left ( {\bk_1\cdot \bk_2 \over k_1 k_2}  \right ) +
{2 \over 7} \left ( {\bk_1\cdot \bk_2 \over k_1 k_2}  \right )^2; \quad k_i = |\bk_i|.
\label{eq:F2}
\een\ees
The perturbative bispectrum as $B^{\rm PT}_{\delta}$ which depends on the linear power spectrum $P^{\rm L}$.
Here, ${\bf k}_i$ represents 3D wave vectors and their moduli are represented as ${k}_i$.
In the flat-sky approximation, the convergence power spectrum $P^{\kappa}$ and bispectrum $B^{\kappa}$ can be expressed
in terms of the underlying power spectrum $P_{\delta}$ and $B_{\delta}$ using the following line-of-sight
integrations:
\bes
\ben
&& P^{\kappa}({l}) = \int_0^{r_s} dr {\omega^2(r) \over d_A^2(r)} 
P_{\delta}\left ({{l} \over d_A(r)}; r \right );\\
&& B^{\kappa}({\bl}_{1},{\bl}_{2},{\bl}_{3}) = \int_0^{r_s} dr {\omega^3(r) \over d_A^4(r)}
B_{\delta}\left ({{\bl}_{1} \over d_A(r)},{{\bl}_{2} \over d_A(r),},
{{\bl}_{3} \over d_A(r)}; r \right).
\een
\ees
In the highly nonlinear regime many different halo-model based fitting functions
have been proposed. We will use the most recent fitting function presented in Ref.\citep{Bihalo}
known to be more accurate compared to the previous fitting functions.
We will use these expressions in our calculation for the Line Correlation Function (LCF)
for the $\kappa$ field. Notice that bispectrum only represents the leading contribution.
Higher-order correction also get contributions from higher-order statistics
such as the trispectrum which we have ignored in our study,

For the validation of our theoretical results we use simulation
that adopted cosmological parameters consistent with the WMAP 9 year result
$\Omega_m=1-\Omega_{\Lambda}=0.279$, $\Omega_{cdm}=0.233$, $\Omega_b=0.046$, $h=0.7$,
$\sigma_8=0.82$ and $n_s=0.97$

%
\section{Third-order Phase Statistics in Projection}
\label{sec:line}
%
%
We will introduce the two-point correlation function
${\xi}(\theta)$ of the weak lensing convergence $\kappa$ through the following expression:
\ben
{\xi}(\theta) = {\langle\kappa(\th_0)\kappa(\th_0+\th)\rangle}
= {1 \over 4\pi} \sum_{2}^{l_{max}} (2l+1){\cal C}_{l} P_{l}(\cos\theta)
\approx {1 \over 2\pi} \sum_{2}^{l_{max}}\, l\, {\cal C}_{l} J_0(l\theta).
\label{eq:two_point}
\een
Here $P_{l}$ is the Legendre polynomial and $l$ represents the angular harmonics.
Also, $J_0$ represents the Bessel Function of order zero of the first kind.
The angular power spectrum ${\cal C}_{l}$ of convergence $\kappa$ is identical to $P_{\kappa}(l)$
for the flat-sky approximation.
We will specialise the discussion to weak lensing convergence $\kappa$ in
the following section, For the purpose of discussion here $\kappa$ is
a generic two-dimensional (2D) field defined over the celestial sphere.
We will consider a flat patch of the sky for our discussion. The
position vector $\th$ is defined using the polar angle $\phi$ and
Cartesian unit vectors ${\bf i}$ and ${\bf j}$:
\ben
{\th} = \theta(\,\hat{\bf i}  \cos\phi + \hat{\bf j} \sin\phi)
\een
Next we will consider two third-order statistics in projection.
\subsection{Line Correlation Function in Projection}
%
%
We will denote the LCF as ${\cal L}_2(\theta)$ which is
defined through the following expression as a function of the angular scale $\theta$ is 
\ben
&& {\cal L}_2(\theta)=
\la \ep(\th_0+\th^{})\ep(\th_0^{}) \ep(\th_0-\th) \ra.
\label{eq:Line}
\een
The LCF is a angle-averaged three-point collapsed correlation function where
the three-points are in a collinear configuration. The two
outer points are equidistant from the central point situated at ${\th_0}$
at a distance $\theta$. $A_s$ represents the survey area.
Assuming isotropy and homogeneity the estimator ${\cal L}(\th)$ depends only on the separation
angular scale $\theta$ and does not depend on the position angle $\th_0$.
We consider the following convention for the Fourier transform relating $\epsilon(\bl)$
and  $\ep(\th)$. Where, $\ep(\th)$ is a real field constructed from the phases $\epsilon(\bl)$
of the convergence map and is constructed from the $\kappa(\bl)$ and its amplitude $|\kappa(\bl)|$, i.e.,
$\epsilon(\bl) = \kappa(\bl)/|\kappa(\bl)|$.
The two-dimensional wave vector is denoted by $\bl$. Then
\ben
&& \ep(\th) = \int {d^2\bl \over (2\pi)^2} \exp(\th\cdot\bl) \epsilon(\bl) W(\bl).
\label{eq:def_epsilon}
\een
The $W(\bl)$ represents the smoothing window. We will not consider observational mask.
The real-space statistics can be estimated simply by avoiding the masked region.
The three-point correlation function of $\epsilon(\th)$ can be expressed as follows:
\ben
\la \epsilon(\bl_1)\epsilon(\bl_2)\epsilon(\bl_3)\ra 
\equiv (2\pi)^2 B^{\epsilon}(\bl_1,\bl_2,\bl_3) \delta_{\rm 2D}(\bl_1+\bl_2+\bl_3).
\een
\begin{figure}
  \begin{center}
  \begin{minipage}[b]{0.45\textwidth}
    \includegraphics[width=\textwidth]{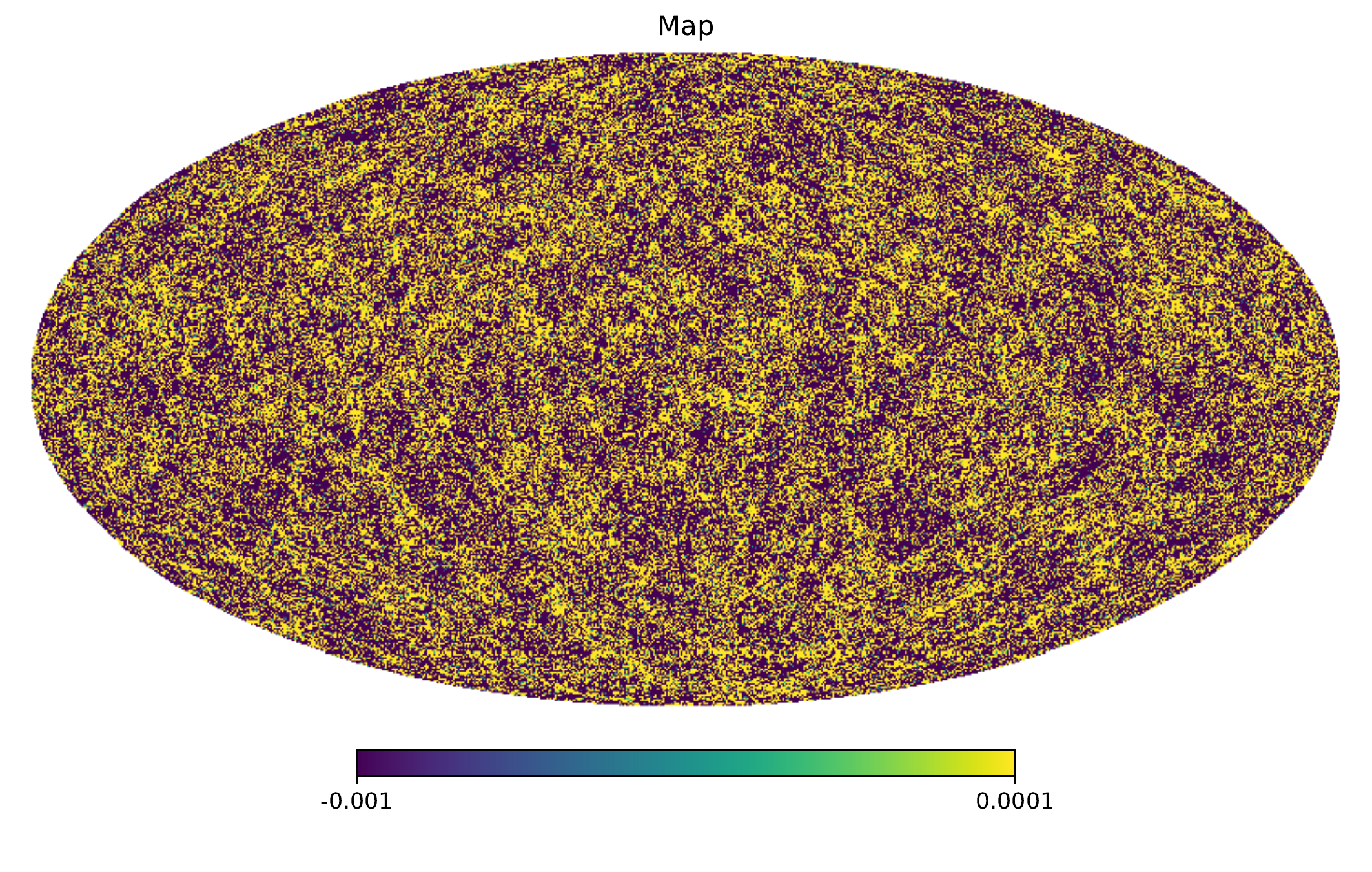}
    \label{fig:2}
  \end{minipage}
  \begin{minipage}[b]{0.45\textwidth}
    \includegraphics[width=\textwidth]{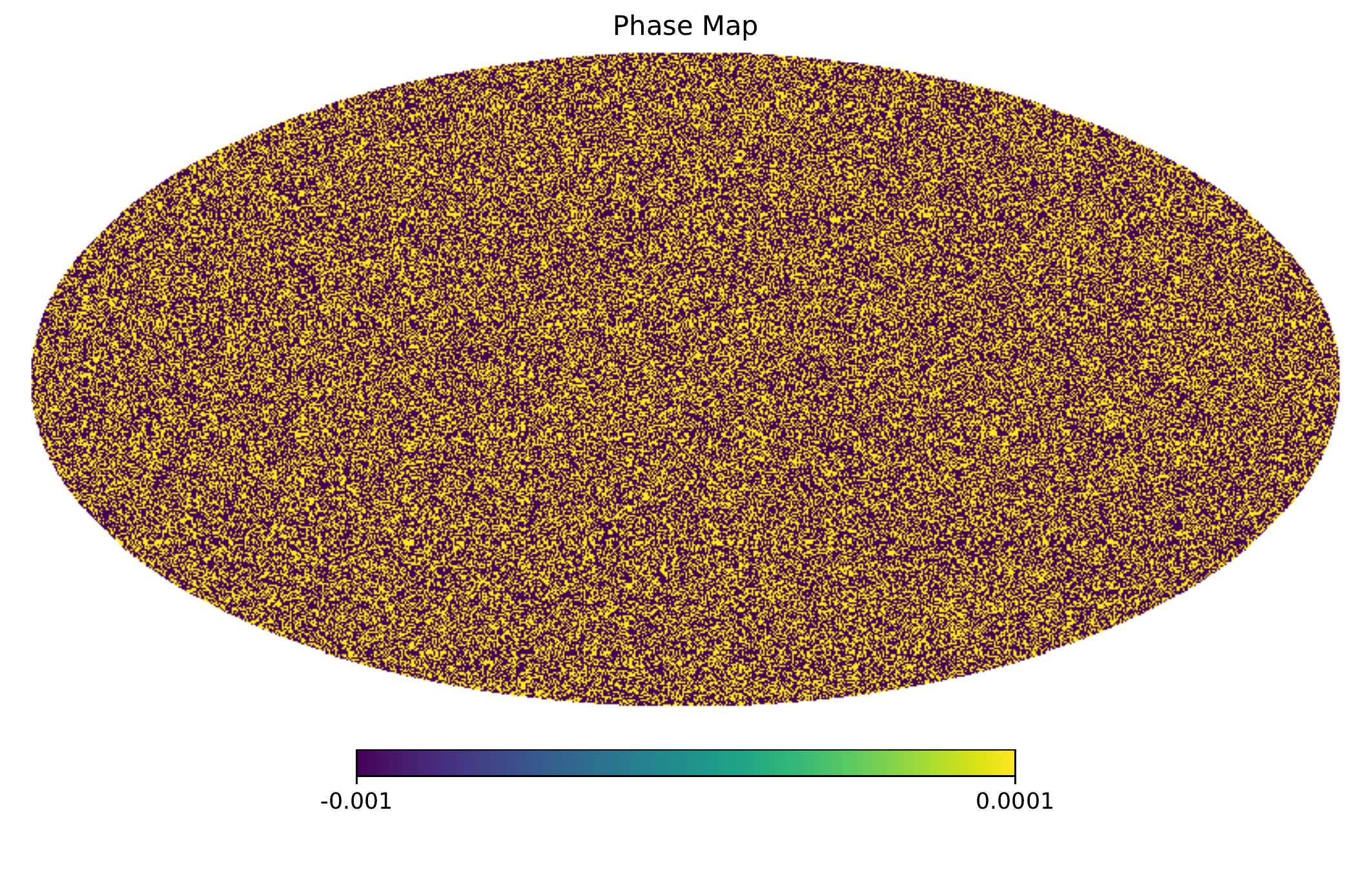}
    \label{fig:1}
  \end{minipage}
  \caption{The left-panel depicts an all-sky  convergence or $\kappa({\bf \theta})$
  map for source redshift $z=0.5$ whereas the right-panel shows
  the corresponding phase $\epsilon({\bf \theta})$ (defined in Eq.({\ref{eq:def_epsilon})}) map for the
  same source redshift. We use the publicly available maps discussed in \citep{Ryuichi}.}
  \label{fig:allsky}
  \end{center}
\end{figure}
The LCF in terms of the bispectrum can be expressed as follows:
\ben
    && {\cal L}_2(\theta) = \int{d\phi \over 2\pi} \int {d^2\bl_1 \over (2\pi)^2}
    \int {d^2\bl_2 \over (2\pi)^2}\int {d^2\bl_3 \over (2\pi)^2}
    \, \la\epsilon(\bl_1)\epsilon(\bl_2)\epsilon(\bl_3)\ra \nn 
    && \quad \times\exp i[\bl_1\cdot\th_0  + \bl_2\cdot(\th+\th_0) + 
    \bl_3\cdot(\th - \th_0)].
    \label{eq:defLine}
\een
The three-point correlation function of the phase can be written in terms of the
convergence bispectrum $B_\kappa(\bl_1,\bl_2,\bl_3)$ 
\ben
\la\epsilon(\bl_1)\epsilon(\bl_2)\epsilon(\bl_3)\ra
= {(2\pi)^2 \over A_s} \left ( {\sqrt \pi\over 2 }  \right )^2
{B_\kappa(\bl_1,\bl_2,\bl_3) \over \sqrt{P_{\kappa}(l_1)P_{\kappa}(l_2)P_{\kappa}(l_3)}}
\delta_{\rm 2D}(\bl_1+\bl_2+\bl_3)
\label{eq:epsilon_bispec}
\een
Using Eq.(\ref{eq:epsilon_bispec}) in Eq.(\ref{eq:Line}) we arrive at the following expression:
\bes\ben
&&    {\cal L}_2(\theta) = \int {d^2 \bl_1 \over (2\pi)^2}\int {d^2 \bl_2 \over (2\pi)^2} {B^\kappa(\bl_1,\bl_2,\bl_1+\bl_2)\over \sqrt{P_{\kappa}(l_1)P_{\kappa}(l_2)P_{\kappa}(|\bl_1+\bl_2|)}} \int {d\phi \over 2\pi} \exp{[i(\bl_1-\bl_2).\th]} \label{eq:corr1} \\
&&    {\cal L}_2(\theta) = \int {d^2 \bl_1 \over (2\pi)^2}\int {d^2 \bl_2 \over (2\pi)^2} {B^\kappa(\bl_1,\bl_2,\bl_1+\bl_2)\over \sqrt{P_{\kappa}(l_1)P_{\kappa}(l_2)P_{\kappa}(|\bl_1+\bl_2|)}} J_0{(|\bl_1-\bl_2|\theta)}.
 \label{eq:corr2}
\een\ees
We have used the following Bessel's first
integral\footnote{\href{http://mathworld.wolfram.com/BesselFunctionoftheFirstKind.html}{\tt Bessel's first integral}} (Eq.(71) of Mathsworld) to reduce Eq.(\ref{eq:corr1}).
\ben
J_n(x) = {1 \over 2\pi\, i^n}\int^{2\pi}_{0}\exp[i(x\cos\tau+in\tau)]d\tau.
\een
Here, $J_n$ denotes the Bessel functions of the first kind of
order $n$.

In addition to the LCF of phases introduced in Eq.(\ref{eq:corr1})-Eq.(\ref{eq:corr2})
we will also consider the following associated estimator:
\bes\ben
&&    {\Pi}_2(\theta) = \int {d^2 \bl_1 \over (2\pi)^2}\int {d^2 \bl_2 \over (2\pi)^2}
B^\kappa(\bl_1,\bl_2,-\bl_1-\bl_2)
\int_0^{2\pi} {d\phi \over 2\pi} \exp{[i(\bl_1-\bl_2).\th]} \label{eq:pi1}. \\
&&    {\Pi}_2(\theta) = \int {d^2 \bl_1 \over (2\pi)^2}\int {d^2 \bl_2 \over (2\pi)^2}
B^\kappa(\bl_1,\bl_2,-\bl_1-\bl_2) J_0{(|\bl_1-\bl_2|\theta)}.
 \label{eq:pi2}
\een\ees
The derivation of Eq.(\ref{eq:pi2}) follows the same steps as
the derivation of Eq.(\ref{eq:corr2}).

\begin{figure}
  \begin{center}
  \begin{minipage}[b]{0.35\textwidth}
    \includegraphics[width=\textwidth]{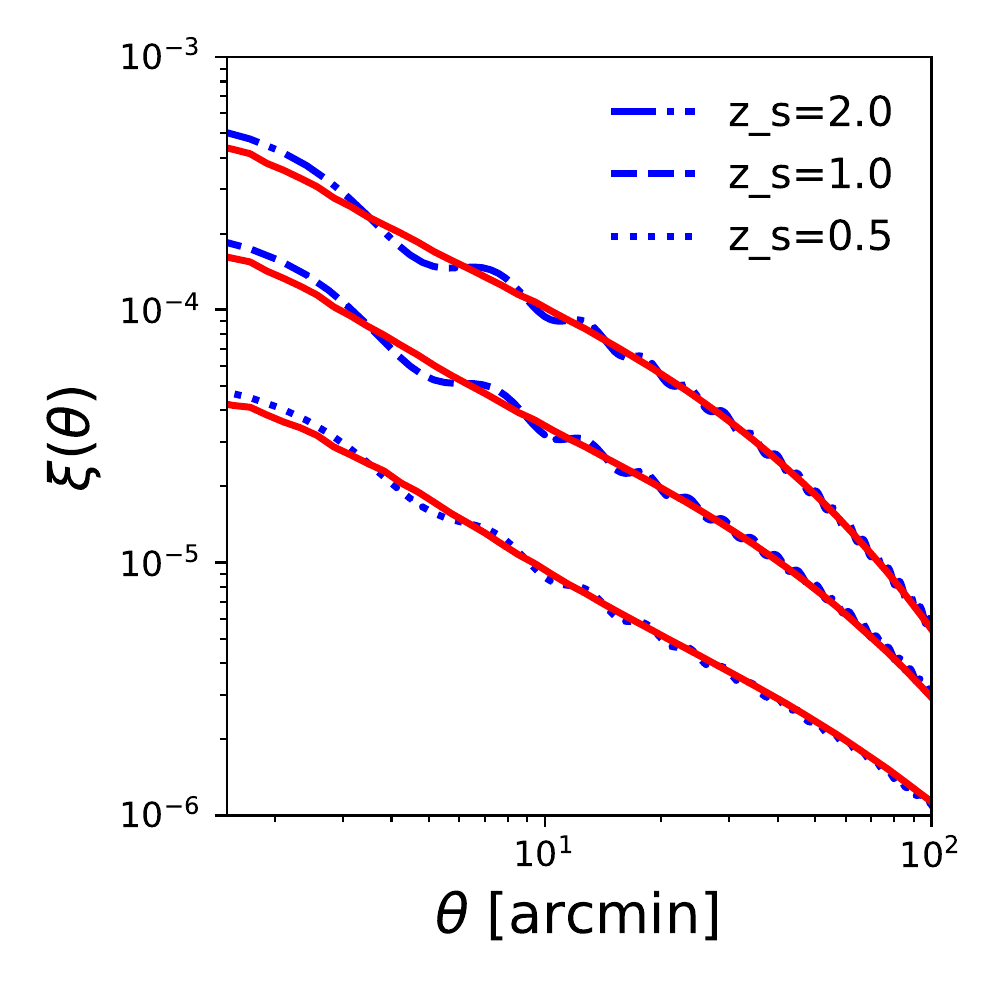}
    \label{fig:1}
  \end{minipage}
  \vspace{-0.5cm}
  \caption{The two-point correlation function $\xi(\theta)$ defined in Eq.(\ref{eq:two_point})
    for the convergence map $\kappa$ is shown as a function of $\theta$. The dot-dashed,
    dashed and dotted lines correspond
    to the theoretical predictions for various redshifts as indicated.
    The solid lines correspond to the estimates from numerical simulations. The maps used
    were degraded to $N_{\rm side}=2048$ before computing the correlation function. \\} 
  \label{fig:2pt}
  %
  %
\end{center}
\end{figure}

\begin{figure}
  \begin{center}
  \begin{minipage}[b]{0.32\textwidth}
    \includegraphics[width=\textwidth]{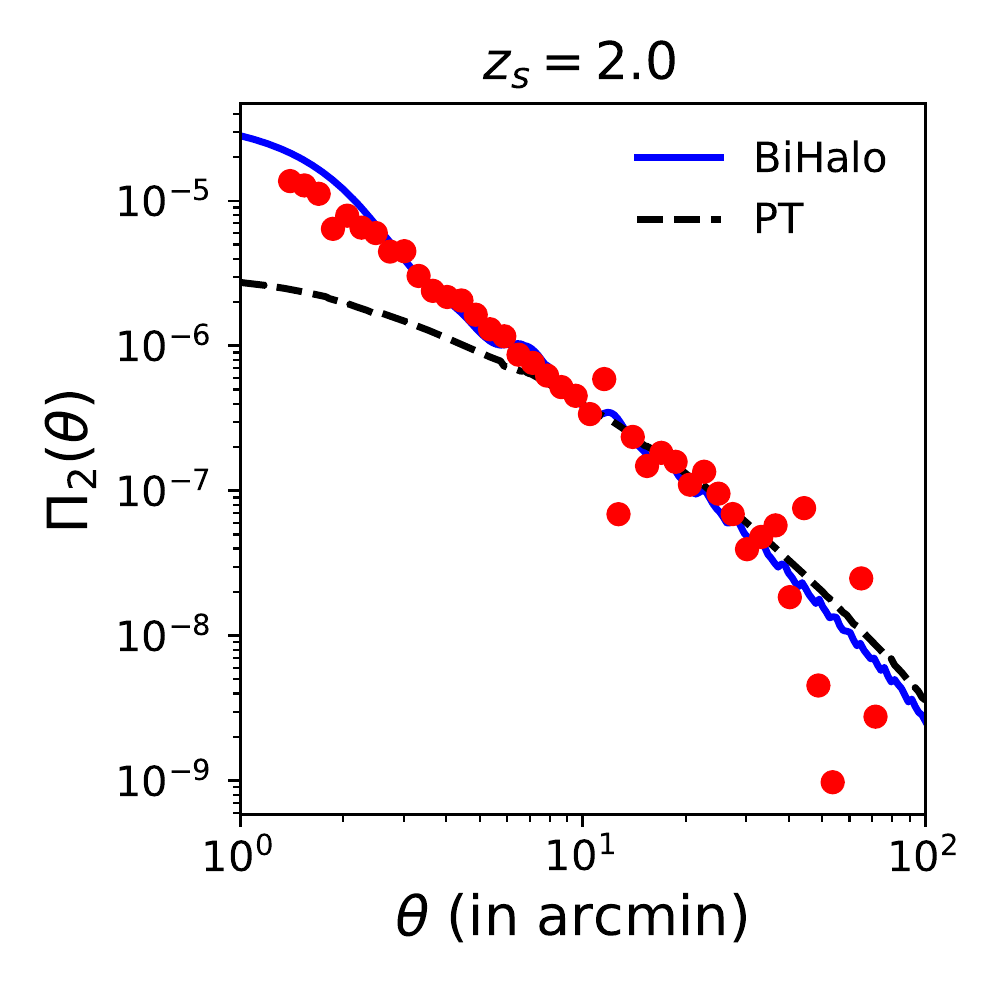}
  \end{minipage}
  \begin{minipage}[b]{0.32\textwidth}
    \includegraphics[width=\textwidth]{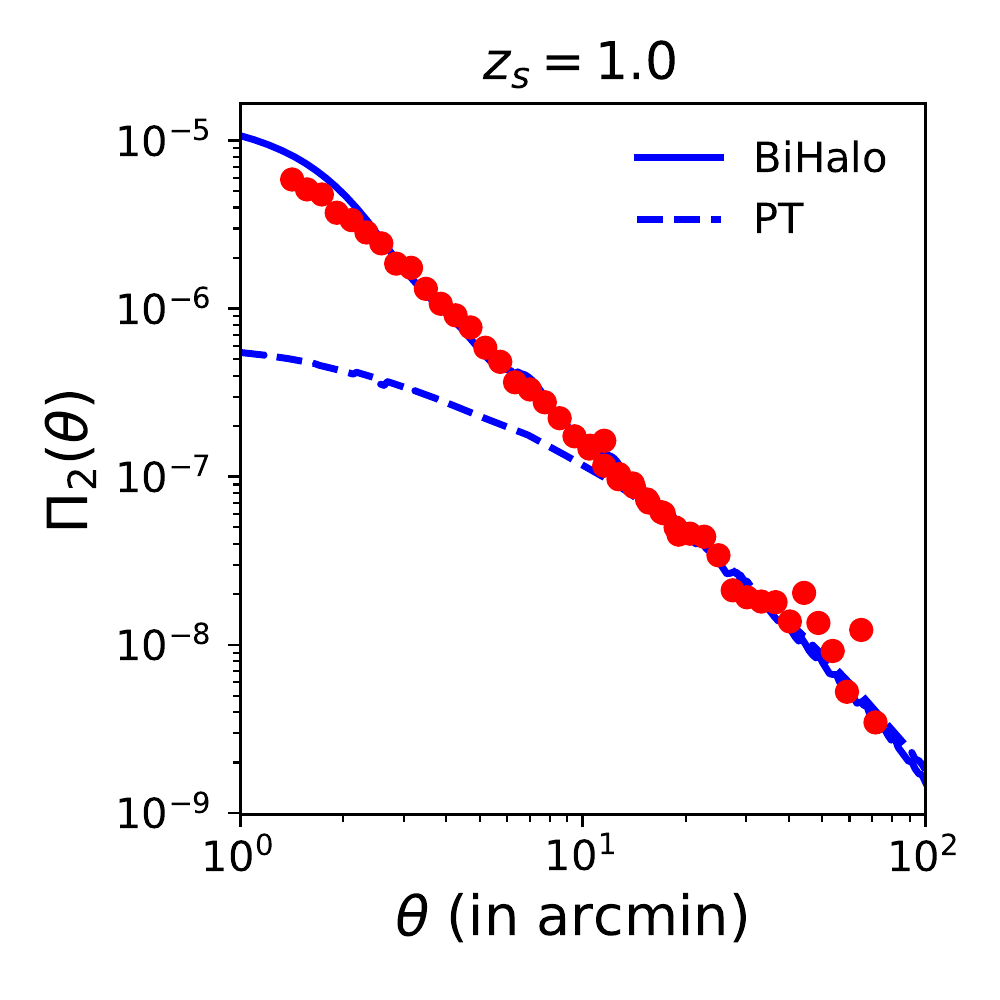}
  \end{minipage}
  \begin{minipage}[b]{0.32\textwidth}
    \includegraphics[width=\textwidth]{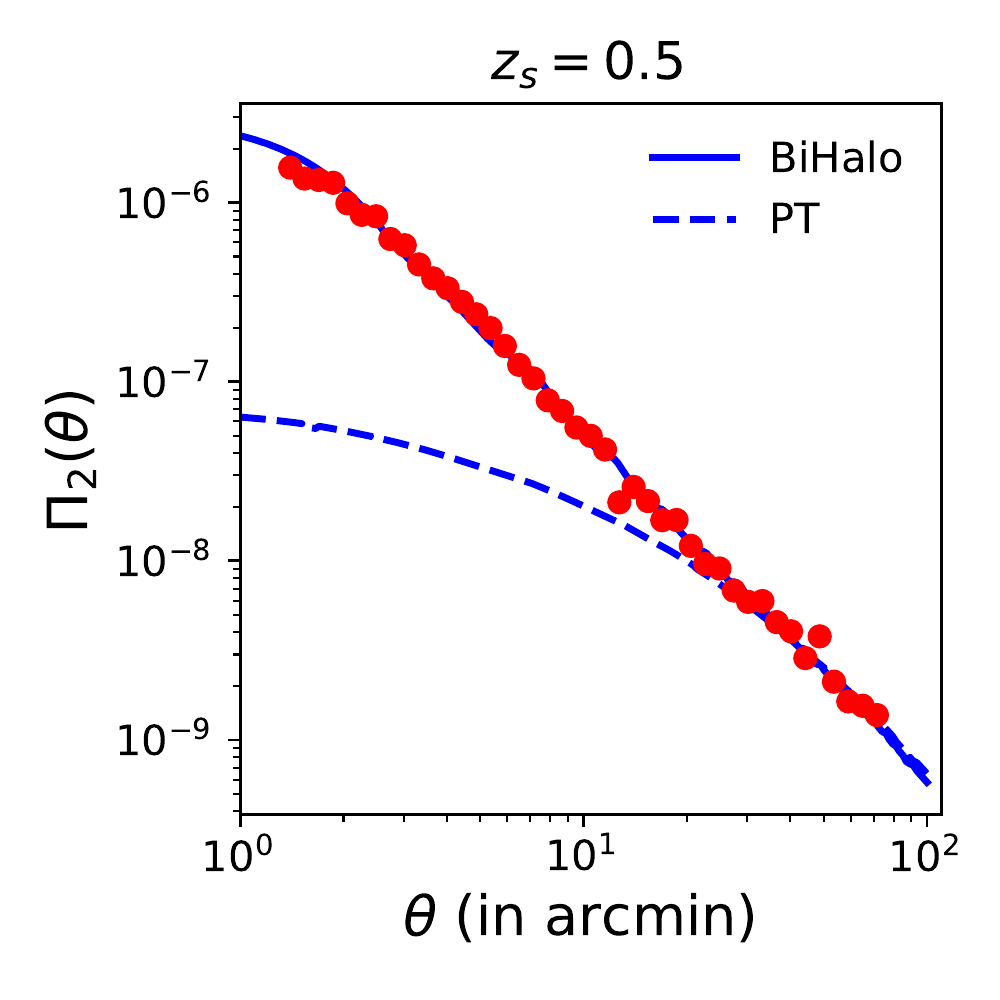}
  \end{minipage}
 \vspace{0.2cm}
 \caption{The line correlation function ${\Pi}_2(\theta)$ is being plotted
   as a function of $\theta$ (in arcmin). The solid lines correspond to
   theoretical predictions defined in Eq.(\ref{eq:pi1})-Eq.(\ref{eq:pi2}).
   We use a fitting function to model the underlying bispectrum.
   The dashed lines are based on perturbation theory (PT) results Eq.\ref{eq:F2}.
   The points are estimates from simulated all-sky
   weak lensing maps. Panels from left to right correspond to
   $z_s=2.0$, $1.0$ and $0.5$ respectively
   We use the fitting function proposed by \citep{Bihalo} for the bispectrum.
   The simulation results show an ensemble
   average computed from 10 all-sky maps  using publicly available software
   {\tt TreeCorr}\citep{{treecorr}}.}
  \label{fig:pi}
  \end{center}
\end{figure}

The Effective Field Theory (EFT) provides a framework to extend the SPT 
results to smaller scales. Including the additional counter-terms
from EFT in the expression of the kernel $F_2$ will extend the validity of
results based on SPT to smaller angular scales not just for gravity
induced non-Gaussianity but also for primordial non-Gaussianity \citep{EFT}.

\subsection{Cumulant Correlators of Phases in Projection}
\label{sec:cc}

The line correlation function considers a collapsed configuration
of a triangle with equidistant points from the central location.
The other collapsed configuration that is commonly used in
the literature was introduced in \citep{{cumu3D}}
in the context of density contrast in $\rm 3D$ and are known
as the cumulant correlators. The two-to-one correlator defined below
is of the lowest-order  in the family of cumulant correlators and can be
constructed cross-correlating a squared $\epsilon(\theta)$ map with itself:

\ben
&& C^{\epsilon}_{21}(\theta) = \la\epsilon^2(\th_0)\epsilon(\th_0+\th)\ra =
    {1\over 4\pi} \sum_{\ell} (2\ell+1) P_{\ell}(\cos\theta) S^{\epsilon}_{\ell}.
    \label{eq:cc}
\een
Here $S^{\epsilon}_{\ell}$ is the skew-spectrum of the $\epsilon(\th)$ map
constructed from its bispectrum $B^{\epsilon}$

\bes
\ben
&& \mathcal{S}^{\epsilon}(l_2) =
\int_0^{\infty} {l_1 d l_1 \over 2 \pi}
\int^{1}_{-1} {d \mu \over 2 \pi \sqrt{1 - \mu^2}}
     {B^{\kappa}(\bl_1,\bl_2,-(\bl_1+\bl_2))\over\sqrt{P_\kappa(l_1)P_\kappa(l_2)P_\kappa(|\bl_1+\bl_2|)}}\\
 \label{eq:2ds0}
 && \mathcal{S}^{\epsilon}_{\ell} = {1\over 4\pi} \sum_{\ell_1\ell_3}
 {B^{\kappa}_{\ell_1\ell_2\ell}\over\sqrt{P_\kappa(l_1)P_\kappa(l_2)P_\kappa}(l)}
 { {(2\ell_1+1)(2\ell_2+1)}  }
\left ( \begin{array} { c c c }
     \ell_1 & \ell_2 & \ell \\
     0 & 0 & 0
\end{array} \right )^2
\een
\ees
Here $\mu$ denotes the cosine of the angle betwwn $\bl_1$ and $\bl_2$ i.e.
$\mu = (\bl_1\cdot \bl_2)/(l_1 l_2)$ with $l_i = |\bl_i|$.
Here the quantity in parentheses is the well-known Wigner-$3j$ symbol.
For more detailed derivation of flat-sky vs. all-sky correspondence
see Appendix-\textsection{\ref{all-vs-flat}}. In Appendix-\textsection{\ref{sec:HigherLCF}}.
we have presented generalisation to higher-order.

The interest in cumulant correlators stems from the fact that they are two-point
correlations but carry information about three-point statistics. They can
be generalised easily to higher order. The information content is in general
different for various triangular configurations. For an equilateral configuration
see \citep{Pritchard}.

\begin{figure}
  \begin{center}
  \begin{minipage}[b]{0.3\textwidth}
    \includegraphics[width=\textwidth]{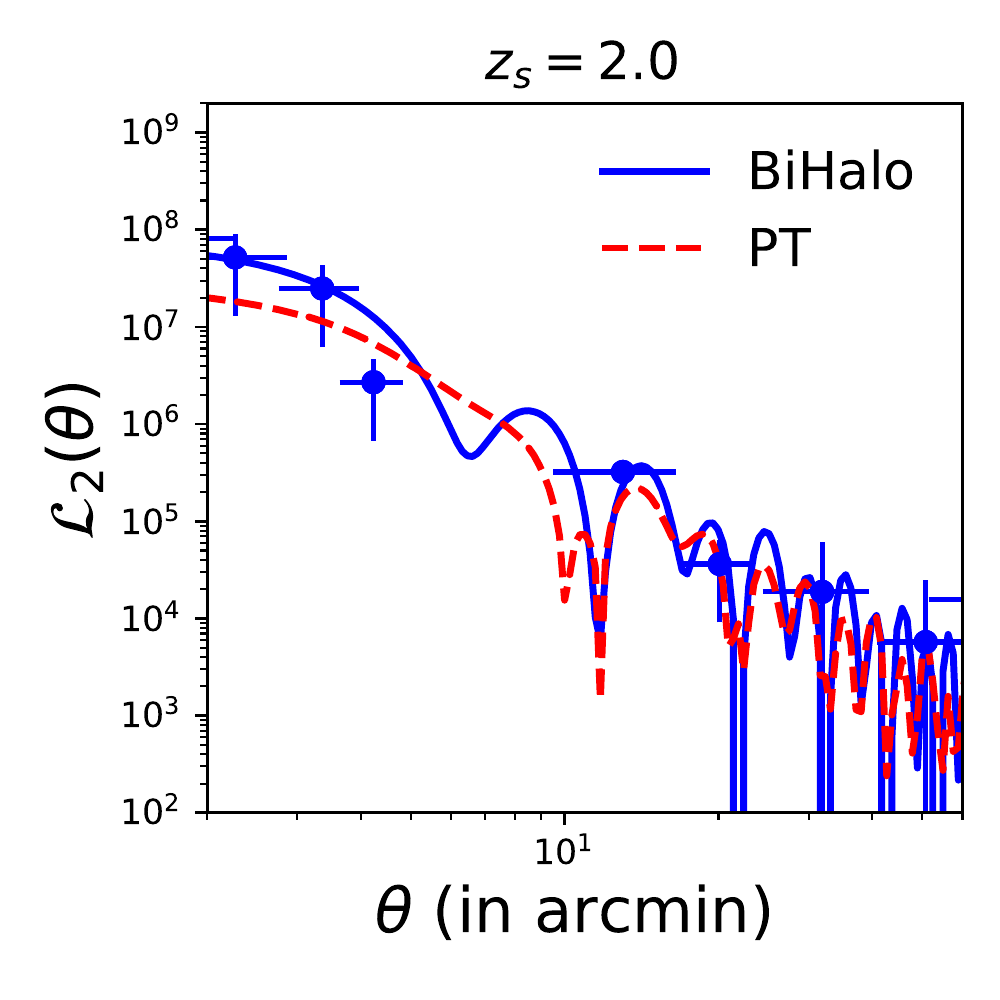}
  \end{minipage}
   \begin{minipage}[b]{0.3\textwidth}
   \includegraphics[width=\textwidth]{Final_corr_zs25.pdf}
   \end{minipage}
   \begin{minipage}[b]{0.3\textwidth}
    \includegraphics[width=\textwidth]{Final_corr_zs25.pdf}
   \end{minipage}
   \label{fig:1}
 \vspace{0.2cm}
 \caption{The line correlation function ${\cal L}_2(\theta)$ defined in
    Eq.(\ref{eq:corr1})-Eq.(\ref{eq:corr2})
   is being plotted as a function of $\theta$ (in arcmin). From left to right we show results for source redshifts
   $z_s=2.0$, $1.0$ and $0.5$. The theoretical predictions correspond to $N_{\rm side}=2048$.
   The dashed-lines correspond to results computed using perturbation theory and the solid lines are
   obtained using a non-perturbative fitting function.
   We use the fitting function proposed by \citep{Bihalo} for the bispectrum.
   The simulation results show an ensemble average computed using 10 all-sky maps. The three-point correlation
   function ${\cal L}_2(\theta)$ was computed from the phase maps using publicly available software
   {\tt TreeCorr}\citep{{treecorr}}.}
  \label{fig:L}
  \end{center}
\end{figure}

\section{Results and Discussion}
\label{sec:disc}
%
In this section we will discuss the results of our numerical investigations for various estimators
and compare them against theoretical predictions.

{\bf Maps:} To validate our analytical results we use the publicly available all-sky weak lensing maps generated by
\citep{Ryuichi}\footnote{http://cosmo.phys.hirosaki-u.ac.jp/takahasi/allsky\_raytracing/}.
The ray-tracing through N-body simulations were used to generate these maps.
These simulation used  $2048^3$ particles to follow the evolution of gravitational clustering.
To generate the convergence $\kappa$ and the corresponding shear $\gamma$ maps multiple lens planes were used;
the source redshifts used were in the range $z_s= 0.05-5.30$.
We have chosen the maps with $z_s=0.5,1.0, 2.0$ for our study.
For CMB maps the lensing potentials were constructed using the
deflection angles which were used to construct the lensing potentials
and eventually the $\kappa$ maps.
These maps include post-Born corrections. Importance of
post-Born terms in lensing statistics in the context of CMB studies were
outlined \citep{LewisPratten}. However, recent studies have shown such corrections are
not important in case of weak lensing statistics which probe lower redshifts.
The convergence maps were generated using an equal area pixelisation scheme
in {\tt HEALPix}\footnote{https://healpix.jpl.nasa.gov/} format\citep{Gorski}.

\begin{figure}
  \begin{center}
  \begin{minipage}[b]{0.3\textwidth}
    \includegraphics[width=\textwidth]{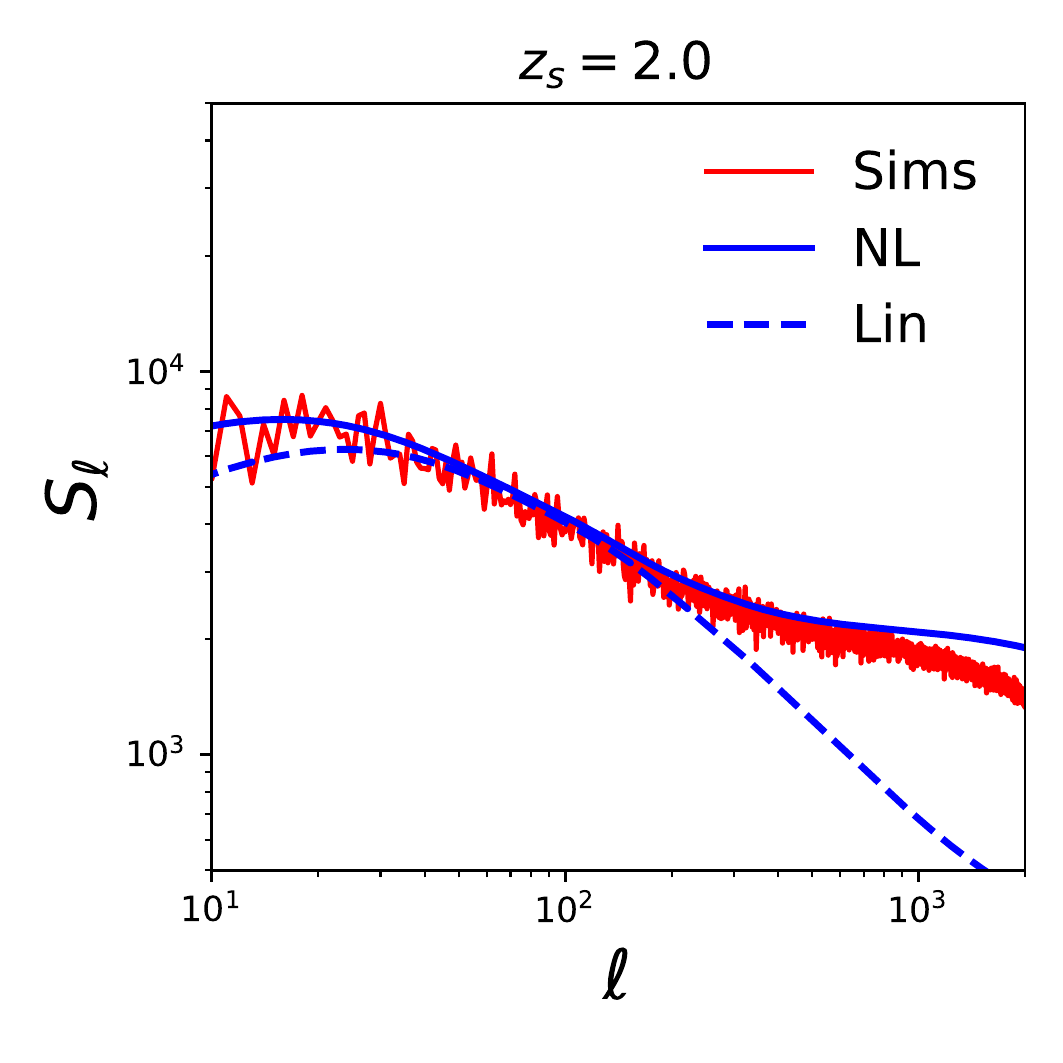}
  \end{minipage}
   \begin{minipage}[b]{0.3\textwidth}
     \includegraphics[width=\textwidth]{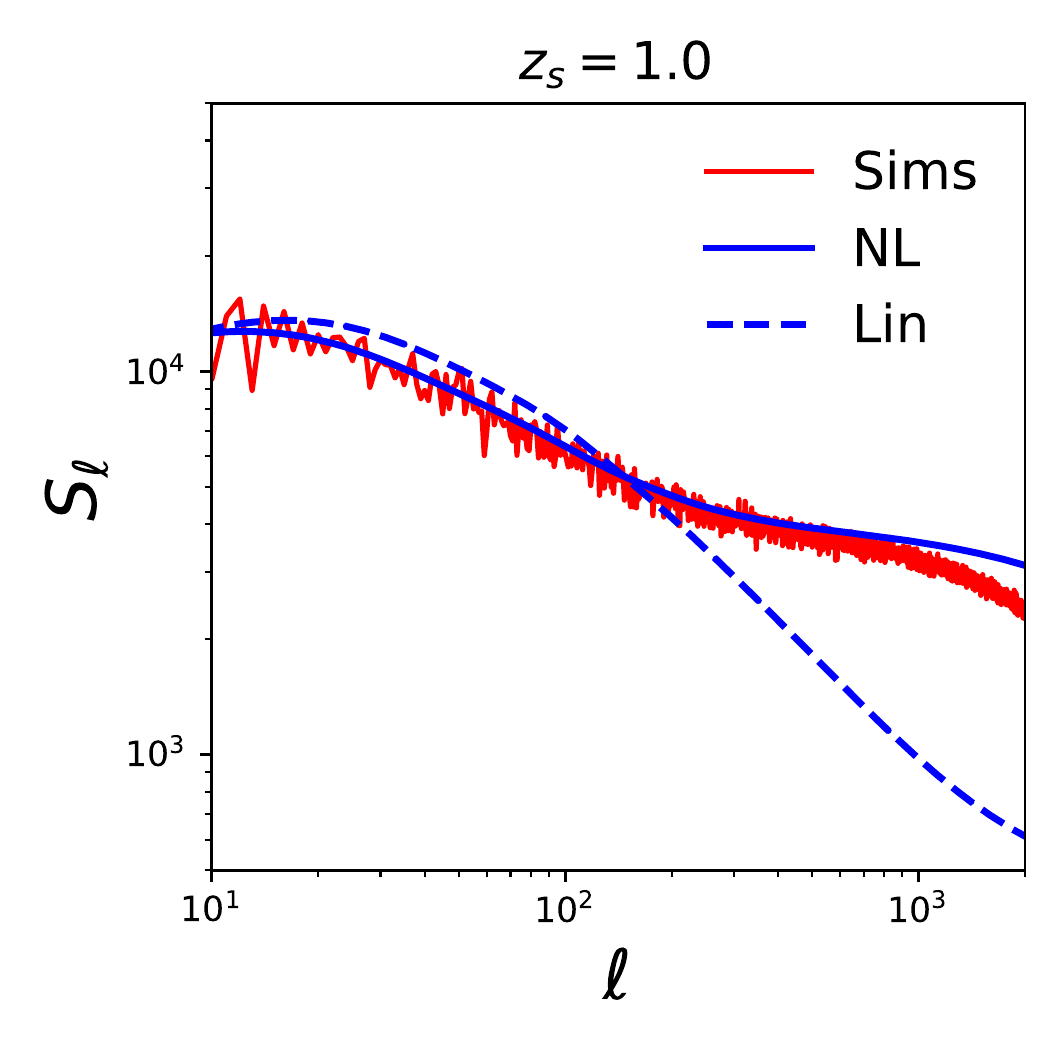}
   \end{minipage}
   \begin{minipage}[b]{0.3\textwidth}
    \includegraphics[width=\textwidth]{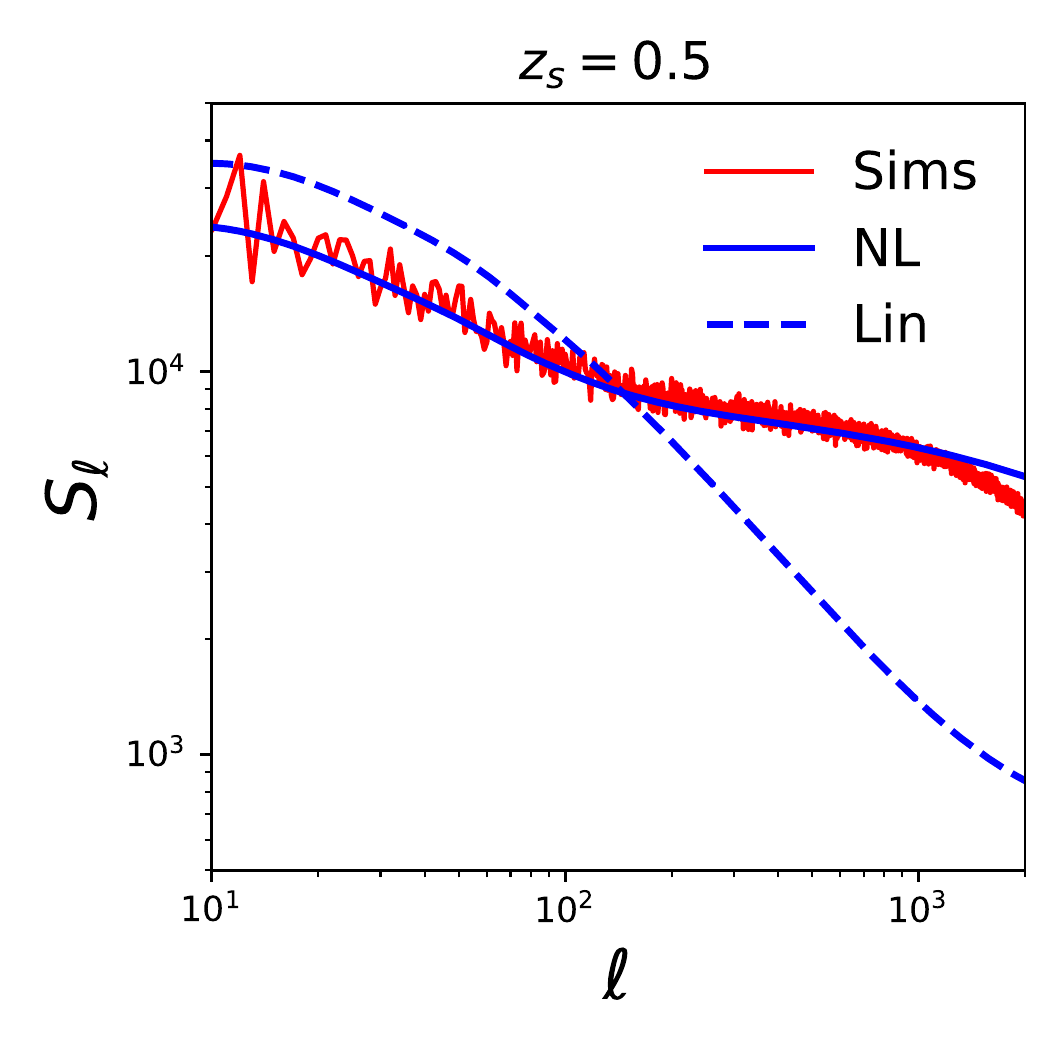}
   \end{minipage}
 \vspace{0.2cm}
 \caption{The solid lines correspond to the skew-spectrum corresponding to the phase maps
   ${S}^{\epsilon}_{\ell}$ defined in Eq.(\ref{eq:2ds0})
   is being plotted as a function of $\theta$ (in arcmin). From left to right we show results for source redshifts
   $z_s=2.0$, $1.0$ and $0.5$. The theoretical predictions correspond to $N_{\rm side}=2048$.
   We use the fitting function proposed by \citep{Bihalo} for the bispectrum. We also show
   the results from perturbation theory (dashed-lines).} 
 \label{fig:skew_plot}
  \end{center}
\end{figure}

The set of maps we use in this study are generated at $N_{\rm side}=4096$
and were cross-checked against higher resolution maps
constructed at a higher resolution $N_{\rm side}=8192, 16384$ for consistency.
These maps constructed at different resolution were found to be
consistent with each other up to the angular harmonics $\ell \le 2000$.

For our study, we have used high resolution maps $N_{\rm side} =4096$.
These maps were degraded to various lower resolution $N_{\rm side} =2048$
In Figure \ref{fig:allsky} we show one such map for the source redshift $z_s=0.5$ (left panel)
and the corresponding phase map (right panel). We analysed these maps
using publicly available software {\tt TreeCorr}\footnote{https://rmjarvis.github.io/TreeCorr/\_build/html/index.html}
for the computation of two- and three-point correlation functions.
The tso-point correlation functions are shown in Figure-\ref{fig:2pt}.

\begin{figure}
  \begin{center}
  \begin{minipage}[b]{0.3\textwidth}
    \includegraphics[width=\textwidth]{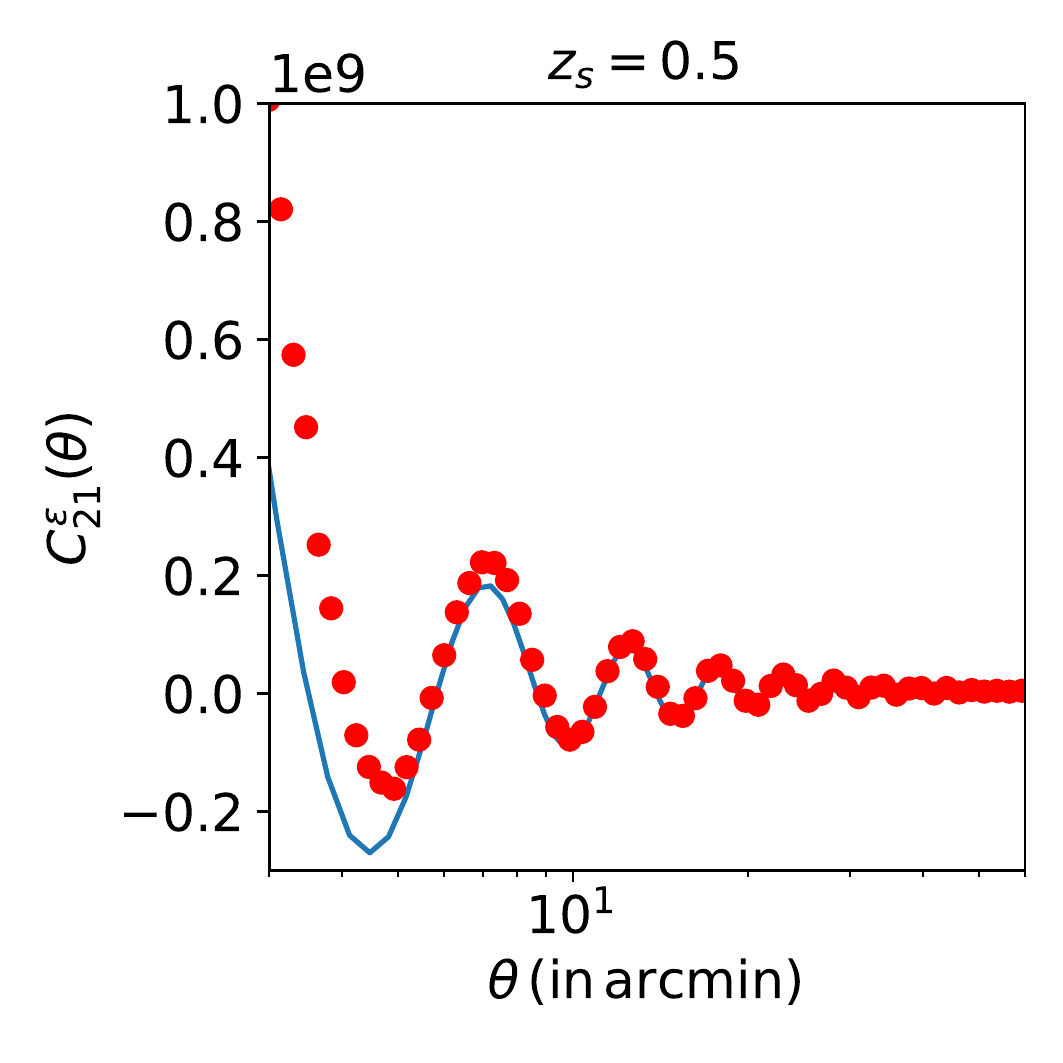}
  \end{minipage}
   \begin{minipage}[b]{0.3\textwidth}
     \includegraphics[width=\textwidth]{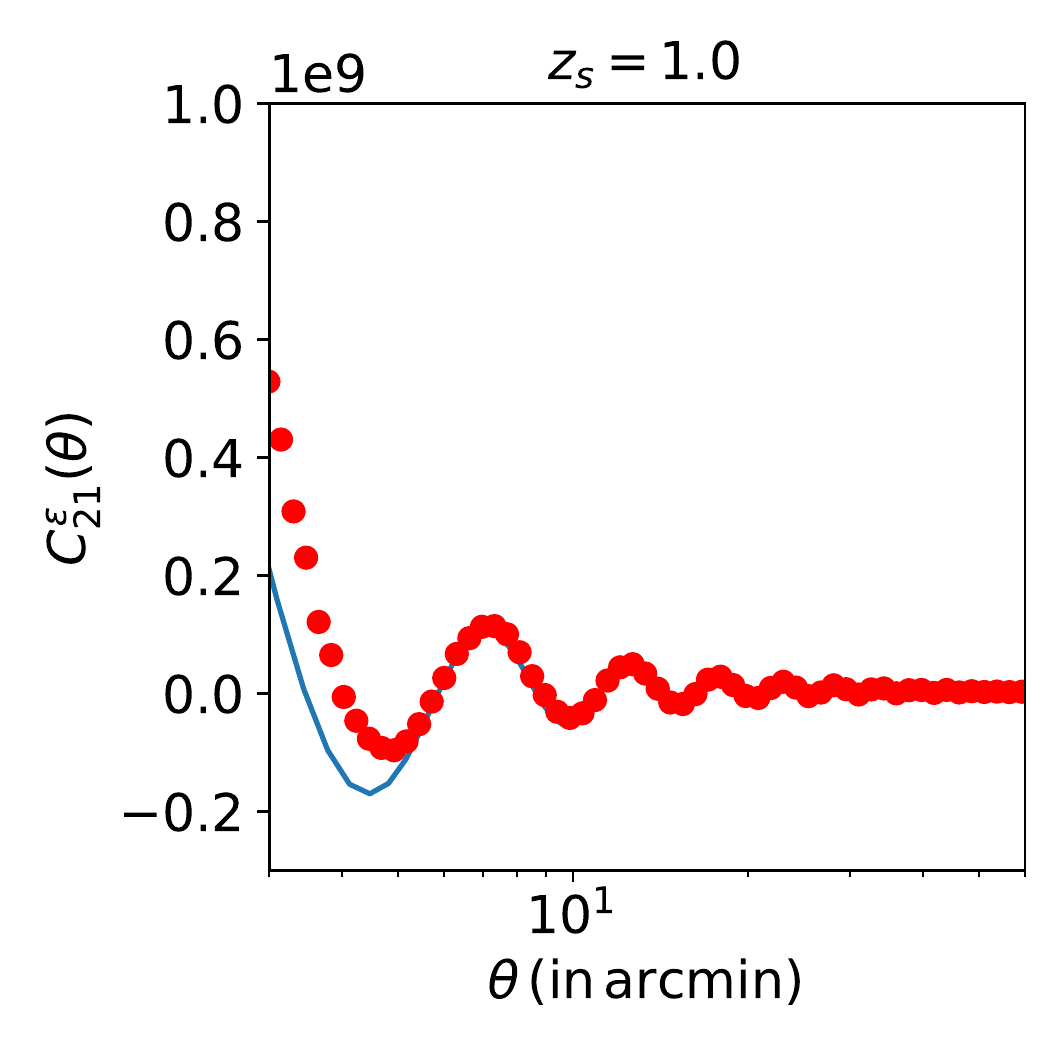}
   \end{minipage}
   \begin{minipage}[b]{0.3\textwidth}
     \includegraphics[width=\textwidth]{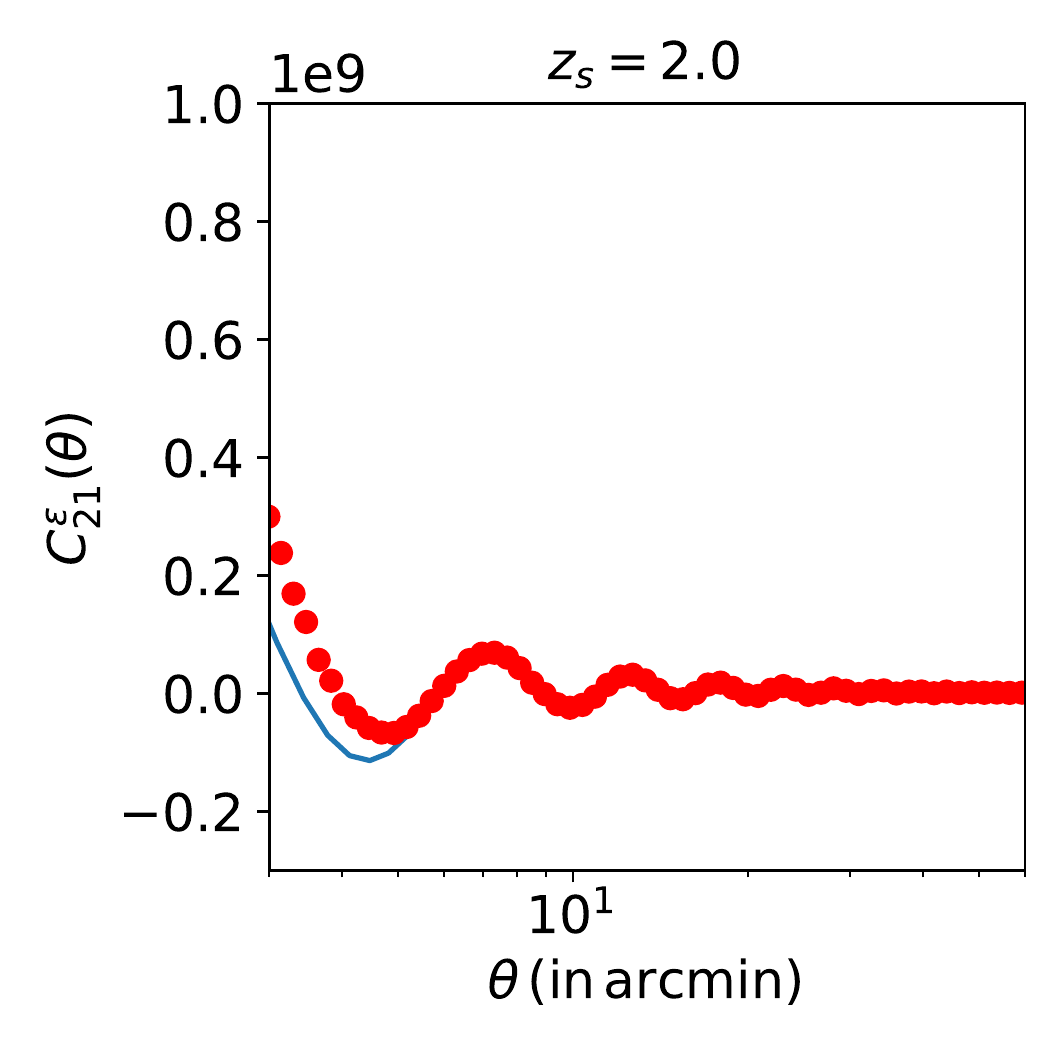}
   \end{minipage}
   \label{fig:cc}
 \vspace{0.2cm}
 \caption{The lines correspond to the cumulant correlator ${C}^{\epsilon}_{21}(\theta)$ defined in Eq.(\ref{eq:cc}).
   is being plotted as a function of $\theta$ (in arcmin). From left to right we show results for source redshifts
   $z_s=2.0$, $1.0$ and $0.5$. The theoretical predictions correspond to $N_{\rm side}=2048$.
   We use the fitting function proposed by \citep{Bihalo} for the bispectrum (solid lines). 
 The dots represent the cumulant correlators defined in Eq.(\ref{eq:cc}) computed 
 from one all-sky phase map by cross-correlating the squared phase map
 $\epsilon^2(\th)$ with itself $\epsilon(\th)$. We have used the publicly available software {\tt TreeCorr}.
 Theoretical predictions correspond to the one computed using the fitting function
 proposed by \citep{Bihalo} for the bispectrum. The deviations from the theoretical predictions at smaller angular
 scales is an effect of pixelisation.} 
 \label{fig:cc}
  \end{center}
\end{figure}

{\bf LCF for $\kappa$:} The line correlation function ${\Pi}_2(\theta)$ defined in Eq.(\ref{eq:corr2})
is being plotted as a function of $\theta$ In Figure - \ref{fig:pi}. From left to right we show results
for source redshifts
   $z_s=2.0$, $1.0$ and $0.5$.  We have presented the mean results estimated from ten all-sky
   maps. No noise was included in our study. Degraded maps with $\rm Nside = 2048$ and $512$ were
   respectively used to estimate the correlation function for the range $\theta_s=1'-10'$
   and $10'-100'$. The fitting function presented in \citep{Bihalo} was used to compute
   the theoretical predictions. For the sake of completeness we have also shown the two-point correlation function
   of the $\kappa$ maps in Figure - \ref{fig:2pt} for various redshifts,
   In this study we will see that at the low source redshift Post-Born correction plays
   a negligible role, but they play a significant role
   at higher redshift, e.g. in case of lensing of CMB.

{\bf LCF for phase:} In Figure - \ref{fig:L} the LCF ${\cal L_2}(\theta)$ is being plotted as a function
of $\theta$ (in arcmin). The solid lines correspond to the ones defined in Eq.(\ref{eq:pi2})
   and the dashed-lines correspond to estimates  from simulated all-sky weak lensing maps.
   Panels from left to right correspond to $z_s=2.0$, $1.0$ and $0.5$ respectively.
   The dashed-lines in each panel correspond to theoretical predictions computed using Born approximation.
   We have checked that the post-Born corrections do not make any appreciable difference.
   We have used ten realisations of all-sky
   maps to compute the numerical estimates. No noise was included.
   Degraded maps with $\rm Nside = 2048$ and $512$ were respectively used to estimate the
   correlation function for the range $\theta_s=1'-10'$
   and $10'-100'$. We use $\ell_{max} = 2N_{\rm side}$ for our study.
   The theoretical results were computed using the fitting function presented in \citep{Bihalo}.
   
\begin{figure}
  \begin{center}
  \begin{minipage}[b]{0.375\textwidth}
    \includegraphics[width=\textwidth]{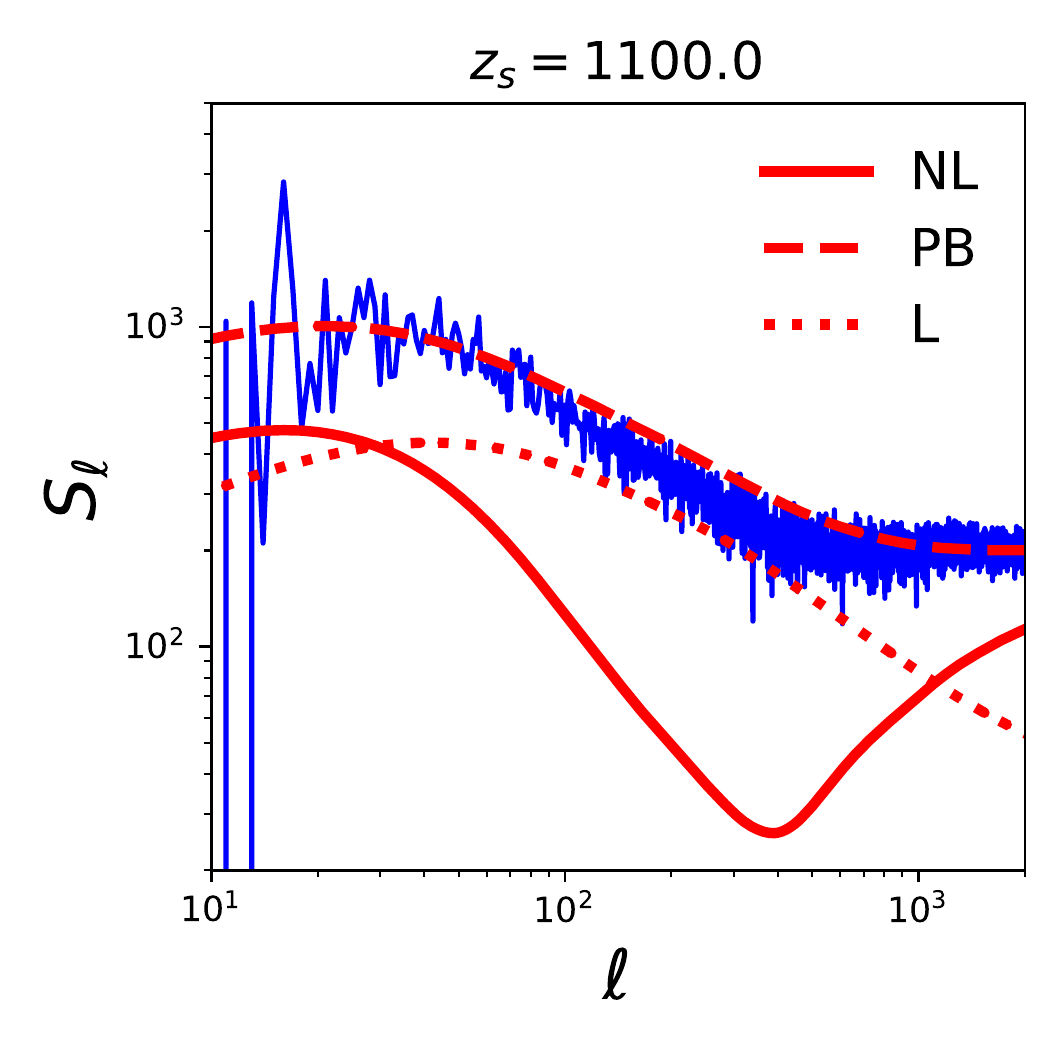}
  \end{minipage}
  \caption{The skew-spectrum $S^{\epsilon}_{\ell}$ corresponding to phase maps $\epsilon(\th)$
    for the source redshift $z_s=1100$
    is plotted as a function angular wave number $\ell$. Three different curves are shown
    Nonlinear (solid-lines), Linear (dotted-lines) and Nonlinear with Post-Born corrections
    included (dashed-lines) along with results from simulations. The simulation results correspond to
    an ensemble average of five independent all-sky simulations. The simulations have $N_{side}=2048$.
  Theoretical results were computed using $\ell_{max}=4096$.}
  \label{fig:skew_cmb}
  \end{center}
\end{figure}

   {\bf Skew-spectrum for phase:} In addition to the LCF we have also introduced
   the skew-spectrum estimator, not for $\kappa(\th)$, but for the phase maps $\epsilon(\th)$
   (denoted as $S^{\epsilon}_{\ell}$) defined in Eq.(\ref{eq:2ds0}).
   For the bispectrum we have used the nonlinear fitting function \citep{Bihalo}
   and for the perturbative calculations we have used the SPT result in Eq.(\ref{eq:F2}).
   We have presented three redshifts $z_s=0.5, 1.0$ and $2.0$ in Fig.\ref{fig:skew_plot} in different panels
   as indicated. The simulation results compare reasonably well against nonlinear predictions.
   We also see departure at high-$\ell$ which is a result of pixelisation. Our simulation results
   are derived using $N_{\rm side}=2048$ which are generated by degrading maps
   originally created at $N_{\rm side}=4096$. For PT results to be valid the maps need to be smoothed.
   We have also included $z_s=1100.0$ in our analysis which is relevant for CMB studies.
   The results are plotted in Fig.\ref{fig:skew_plot} and for CMB lensing
   in Fig.\ref{fig:skew_cmb}. We found that even with a single all-sky realisation we can estimate the
   $S^{\epsilon}_{\ell}$ with a very high degree of accuracy. We have included realistic noise though
   it is expected from our previous study \citep{skew} inclusion of noise will not change our findings completely.

   The importance of Post-Born (PB) corrections for CMB lensing lensing,
   was underlined in many recent studies, e.g., Ref.\citep{skew}. While such corrections 
   do not make any significant contribution they do play important role for high redshift CMB studies.
   In Fig.\ref{fig:skew_cmb} we have
   shown the nonlinear results with and without PB corrections. Inclusion of Post-Born
   corrections is important to reproduce the simulation results.
   The maps used were of resolution $N_{\rm side}=2048$.
   Higher-order generalisations of ${\cal L}_2$ is presented in \textsection\ref{sec:HigherLCF}.
   
   {\bf Cumulant correlators for phase:} The cumulant correlators (CCs) carry equivalent information
   compared to the skew-spectrum $S^{\epsilon}_{\ell}$ as they can be constructed from skew-spectrum.
   The theoretical cumulant correlator $C^{\epsilon}_{21}(\theta)$ using Eq.(\ref{eq:cc}).
   The numerical results were computed using the {\tt TreeCorr}.
   The results are shown in Fig.\ref{fig:cc}. 
   This alows computation of CCs without broad binning as was the case for LCF.
   The deviation from theoretical prediction in the low-$\theta$ regime is related to the
   pixelisation effect. A comparison with the results presented for ${\cal L}_2$ in Fig.\ref{fig:L}
   confirms very high S/N in $C^{\epsilon}_{21}(\theta)$. The generalisation to higher-order
   is straightforward and can by cross-correlating $p$-th power of $\epsilon(\th)$,i.e., $\epsilon^p(\th)$
   against its $q$-th power, i.e., $\langle\epsilon^q(\th_1)\epsilon^q(\th_2)\rangle$.

\section{Conclusions and Future Prospects}
\label{sec:conclu}

The primary aim of this paper was to introduce statistics of
Fourier phases in the context of weak lensing studies.
We have introduced the three-point phase correlation LCF
to probe the non-Gaussianity in weak lensing convergence maps used
in the study of galaxy clustering ${\cal L}_2$.
In addition, an associated three-point statistics ${\Pi}_2$ was also introduced for
probing statistics of $\kappa$ maps. We have used a set of state-of-the-art 
all-sky simulations of weak lensing convergence maps to test
our theoretical results as a function of source redshift.
We have generalised both ${\Pi}_2(\theta)$ and ${\cal L}_2(\theta)$ to higher-order.
Next, we have adopted the cumulant correlator typically used for 3D density field
and 2D convergence maps for statistics phases. We showed that available theoretical
models can reproduce the numerical results with reasonable accuracy and our
results can be used to select the range to retain in order to maintain a given level
of theoretical accuracy. While we have focussed on weak lensing convergence, the statistical estimators presented
here will be useful in other areas cosmology, e.g., galaxy clustering (in real and redshift space)
and in clustering of Lyman-$\alpha$ absorbers.

Several extensions are possible on the theoretical
front. The Effective Field Theory (EFT) provides a framework to extend the validility domain
of the standard perturbation theory (SPT).  A formulation of the phase statistics
in EFT will improve its domain of validity.
It will also be interesting to formulate the phase-statistics directly for shear.
Theoretical modelling of bispectrum generated by intrinsic alignment \citep{Vlah} will also be useful.
The perturbative regime is not particularly affected by gas physics.
Nevertheless, $k$-cut filtering can
be included \citep{BNT} to filter out particularly sensitive modes.

Finally, our study was performed in a rather idealized observational setting to establish
a baseline. Realistically complex follow-up studies will
be presented in future. In particular a Fisher-based analysis independently and
jointly with power spectrum will be presented elsewhere.

\section*{Acknowledgment}
DM was supported by a grant from the Leverhulme Trust at MSSL when this work was initiated.
RT is spported by MEXT/JSPS KAKENHI Grant Numbers 20H0 5855, 20H0 4723 and 17H0 1131.
DM would like to thank Peter Taylor for many useful discussions
regarding implementation of the software {\tt TreeCorr}.
We would like to thank Alexander Eggemeier for many useful discussions.

\appendix

\section{All-sky Expressions}
\label{all-vs-flat}

Our primary aim in this appendix is to develop all-sky estimators presented in the text of the paper.
Following \citep{Martin, ZS, Hu} the spherical polar co-ordinate $(\theta,\phi)$, radial co-ordinate $r = 2 \sin \theta/2 \approx \theta$
Equivalently in the harmonic domain $|\bl|=\ell$ and $\phi_{\bl}$ denotes the polar angle:

\ben
\kappa(\th) = {1\over (2\pi)^2} \int \kappa(\bl)\exp(i\bl\cdot\th) d^2\bl \approx \sum_{\ell m}\kappa_{\ell m}Y_{\ell m}(\th)
\een
The flat-sky $\kappa(\bl)$ and its all-sky harmonic counterpart $\kappa_{\ell m}$ are
related by the following expression:
\bes\ben
&& \kappa(\bl) = \sqrt{4\pi \over 2\ell+1 } \sum_m i^m \kappa_{\ell m}\exp(im\phi_\bl) \\
&& \kappa_{\ell m} = \sqrt{2\ell+1 \over 4\pi} \int {d\phi_\bl \over 2\pi}\exp(-im\phi_\bl) \kappa(\bl)
\een\ees

The spherical harmonic basis $Y_{\ell m}(\theta,\phi)$ and rhe flat sky basis are
related by the following expressions:
\bes
\ben
&& Y_{\ell m}(\theta,\phi) = (-1)^m \sqrt{(2\ell +1)(l-m)! \over 4\pi (\ell+m)!}
P^m_{\ell}(\cos\theta) \exp(im\phi)  \label{eq:YLM}\\
&& P_{\ell}^{-m}(\cos \theta) = (-1)^m {(l-m)! \over (l+m)!} P^m_{\ell}(\cos\theta)
\label{eq:PLM}
\een
\ees
Using the expressions Eq.(\ref{eq:PLM}) in Eq.(\ref{eq:YLM})
we obtain the flat-sky counterpart of the spherical harmonics:
\bes
\ben
&& Y_{\ell m}(\theta,\phi) =  J_{m}(\ell\theta)\sqrt{\ell \over 2\pi}\exp(im\phi).
\een
If we use the following expansion of the plane wave:
\ben
&& \exp(i\bl\cdot\th) = \sum_m i^m J_{m}(\ell\theta)\exp[im(\phi-\phi_{\bl})]  \nn 
&& \quad\quad \approx \sqrt{2\pi \over \ell }\sum_m i^m Y_{\ell m}(\theta) \exp(im\phi_{\bl}).
\een
\ees

Next, we consider the case of bispectrum and related estimator:
\ben
&& \la \kappa_{\ell_1 m_1}\kappa_{\ell_2 m_2}\kappa_{\ell_3m_3}\ra_c \equiv 
B^\kappa_{\ell_1\ell_2\ell_3}
\left ( \begin{array} { c c c }
     \ell_1 & \ell_2 & \ell_3 \\
     m_1 & m_2 & m_3
\end{array} \right ).
\een

\noindent
Here the quantity in parentheses is the well-known Wigner-$3j$ symbol which
enforces the rotational invariance. It is only non-zero for the triplets
$(\ell_1,\ell_2,\ell_3)$ that satisfy the {\em triangular condition} and
$\ell_1+\ell_2+\ell_3$ is even.
The reduced bispectrum $b^\kappa_{\ell_1\ell_2\ell_3}$ is useful in directly
linking the all-sky bispectrum and its flat-sky counterpart.
For the convergence field $\kappa$, $b^\kappa_{\ell_1\ell_2\ell_3}$
is defined through the following expression: 
\ben
&& B^\kappa_{\ell_1\ell_2\ell_3} := \sqrt{(2\ell_1+1)(2\ell_2+1)(2\ell_3+1)\over 4\pi}
\left ( \begin{array} { c c c }
     \ell_1 & \ell_2 & \ell_3 \\
     0 & 0 & 0
\end{array} \right ) b^\kappa_{\ell_1\ell_2\ell_3}.
\een
The flat-sky bispectrum is similarly defined through:
\ben
&& \langle \kappa(\bl_1)\kappa(\bl_2)\kappa(\bl_3)\rangle_c =
(2\pi)^2 \delta_{2D}(\bl_1+\bl_2+\bl_3) B^\kappa(\bl_1,\bl_2,\bl_3). 
\een
The flat-sky bispectrum $B^\kappa(\bl_1,\bl_2,\bl_3)$ is identical to the reduced bispectrum
$b^{\kappa}_{\ell_1\ell_2\ell_2}$ for high multipole \citep{review_ng}. This can be shown
by using the following asymptotic relationship:
\bes\ben
&& {\cal G}_{\ell_1m_1,\ell_2m_2,\ell_3m_3} \equiv \int d\oh Y_{\ell_1m_1}(\oh)
Y_{\ell_2m_2}(\oh)Y_{\ell_3m_3}(\oh) \nn 
&&  = \sqrt {   {(2\ell_1+1)(2\ell_2+1)( 2\ell+1 ) \over 4\pi}  }
\left ( \begin{array} { c c c }
     \ell_1 & \ell_2 & \ell \\
     0 & 0 & 0
  \end{array} \right )\left ( \begin{array} { c c c }
     \ell_1 & \ell_2 & \ell \\
     m_1 & m_2 & m_3
\end{array} \right ) \nn 
&& \approx (2\pi)^2\delta_{\rm 2D}(\bl_1+\bl_2+\bl_3).
\een\ees
The skew-spectrum in the flat-sky is given by \citep{MunshiPratten}:
\bes
\ben
&& \mathcal{S}(l_2) =
\int_0^{\infty} {l_1 d l_1 \over 2 \pi}
\int^{1}_{-1} {d \mu \over 2 \pi \sqrt{1 - \mu^2}}
B^{\kappa}(\bl_1,\bl_2,-(\bl_1+\bl_2)) \nn 
&& \quad\quad\beta(l_1\theta_s)\beta(l_2\theta_s) \beta(| {\bf{l}}_1 + {\bf{l}}_2 |\theta_s).\\
 \label{eq:2ds0}
 && \mathcal{S}_{\ell} = \sum_{\ell_1\ell_3} b^{\kappa}_{\ell_1\ell_2\ell}.
 {   {(2\ell_1+1)(2\ell_2+1)\over 4\pi}  }
\left ( \begin{array} { c c c }
     \ell_1 & \ell_2 & \ell \\
     0 & 0 & 0
  \end{array} \right )^2 \beta_{\ell_1}(\theta_s) \beta_{\ell_2}(\theta_s)  \beta_{\ell}(\theta_s).
\een
\ees
Here the $\beta$ functions represent the smoothing window functions and $\mu$ is the
cosine of the angle between vector $\bl_1$ and $\bl_2$ i.e. $\mu = (\bl_1\cdot\bl_2)/(l_1l_2)$, where $|\bl_i|=l_i$.
The skewness in terms of these expressions is given by:
\ben
\int_0^{\infty} {dl\over 2\pi} S^{\epsilon}({l}) = \sum_{\ell}(2\ell+1)S_{\ell}.
\een

\section{Higher-order LCFs}
\label{sec:HigherLCF}
%
In this section we consider generalisation of the third-order LCFs to
higher-order. The family of LCFs can be seen as a natural extension
of two-point cumulant correlators often used in the literature.
LCFs can be thought of as three-point generalisation of two-point
cumulant correlators. The following statistics are relevant for
the kappa field as well as the for the phases.
In this section we generalise the line-correlation to higher order. The fourth- and
fifth-order generalisations are as follows.
\bes
\ben
&& {\cal L}_4(\theta) \equiv
\int_0^{2\pi}{d\phi \over 2\pi}\la \epsilon(\th_0+\th)\epsilon^2(\th_0)\epsilon(\th_0-\th) \ra  \nonumber \\
&& = \int {d^2 \bl_1 \over (2\pi)^2}\cdots\int {d^2 \bl_3 \over (2\pi)^2} {B_{(4)}^{\kappa}(\bl_1,\bl_2,\bl_3, -\bl_1-\bl_2-\bl_3)\over
  \sqrt{P_{\kappa}(l_1)\dots
    P_{\kappa}(|\bl_1+\bl_2+\bl_3|)}} J_0(|\bl_2-\bl_3|\theta) \label{eq:corr4} 
\een
We have omitted the numerical prefactor.
Equivalently we can generalise Eq.(\ref{eq:pi2}) to fourth-order:
\ben
&& {\Pi}_4(\theta) \equiv \int_0^{2\pi}{d\phi \over 2\pi}\la \kappa(\th_0+\th)\kappa^2(\th_0)\kappa(\th_0-\th) \ra  \nonumber \\
&& = \int {d^2 \bl_1 \over (2\pi)^2}\cdots\int {d^2 \bl_3 \over (2\pi)^2}
B_{(4)}^{\kappa}(\bl_1,\bl_2,\bl_3, -\bl_1-\bl_2-\bl_3) J_0(|\bl_2-\bl_3|\theta).
\een
\ees
Here, $B^{(4)}_{\kappa}$ is the trispectrum for the convergence field and is
defined as follows:
\ben
\langle \kappa(\bl_1)\cdots\kappa(\bl_n) \rangle
= (2\pi)^2\delta_{\rm 2D}(\bl_1+\cdots+\bl_n)B^{(n)}_{\kappa}(\bl_1,\cdots,\bl_n)
\een
The extension to higher-order can be done in a straight-forward manner:
\ben
&&  {\cal L}_5(\theta) \equiv \int_0^{2\pi}{d\phi \over 2\pi} \la \epsilon(\th_0+\th)\epsilon^3(\th_0)\epsilon(\th_0-\th) \ra \nonumber \\
&& = \int {d^2 \bl_1 \over (2\pi)^2}\cdots\int {d^2 \bl_4 \over (2\pi)^2}
    {B_{(5)}^\kappa(\bl_1, \cdots,\bl_4, -\bl_1 \cdots -\bl_4)\over
      \sqrt{P_{\kappa}(l_1)\cdots P_{\kappa}(|\bl_1+\cdots+\bl_4|)}} J_0(|\bl_3-\bl_4|\theta)  \label{eq:corr5} 
\een
There are no numerical fitting functions beyond bispectrum. Nevertheless, the possibility
of upcoming experiments to detect the higher-order correlation has fuelled 
development in this area in recent years. Notice that these estimators
can be useful in diectly probing galaxy clustering in real \citep{Alex} as well
as in redshift space \citep{Obreschkow2} with suitable modifications.

\end{document}